# Interstate Vibronic Coupling Constants Between Electronic Excited States for Complex Molecules

Maria Fumanal,[a] Felix Plasser,[b] Sebastian Mai,[b] Chantal Daniel,[a] and Etienne Gindensperger[a*]

[a] *Laboratoire de Chimie Quantique, Institut de Chimie Strasbourg, UMR-7177 CNRS/Université de Strasbourg, 1 Rue Blaise Pascal BP 296/R8, F-67008 Strasbourg, France.*

[b] *Institute of Theoretical Chemistry, Faculty of Chemistry, University of Vienna, Währinger Straße 17, 1090 Vienna, Austria.*

## Abstract

In the construction of diabatic vibronic Hamiltonians for quantum dynamics in the excited-state manifold of molecules, the coupling constants are often extracted solely from information on the excited-state energies. Here, a new protocol is applied to get access to the interstate vibronic coupling constants at the time-dependent density functional theory level through the overlap integrals between excited-state adiabatic auxiliary wavefunctions. We discuss the advantages of such method and its potential for future applications to address complex systems, in particular those where multiple electronic states are energetically closely lying and interact. As examples, we apply the protocol to the study of prototype rhenium carbonyl complexes $[Re(CO)_3(N,N)(L)]^{n+}$ for which non-adiabatic quantum dynamics within the linear vibronic coupling model and including spin-orbit coupling have been reported recently.

## INTRODUCTION

Excited-state non-adiabatic quantum dynamic simulations for transition metal complexes are very challenging for current state-of-the-art methods, due to specificities of this class of molecules: i) high density of electronic excited states; ii) multiple relevant spin multiplicities; and often iii) high nuclear dimensionality; and iv) low symmetry. For this purpose the linear vibronic coupling model[1,2,3,4] (LVC) has been augmented to include spin-orbit coupling (SOC) and was recently used for describing ultrafast intersystem crossing (ISC) processes driven by spin-vibronic mechanism in various complexes.[5,6,7,8] In such LVC models, the potential energy surfaces of all considered states are constructed from some potential for the ground state, augmented by constant and linear terms in the diabatic potential energy matrix. In the investigations of these complexes, these terms include intrastate and interstate coupling as well as SOC, in order to take into account interactions within and between different spin multiplicities. The results of the simulations reproduce well the time-scales of the luminescent decays observed experimentally for a series of Re(I) α-diimine carbonyl complexes,[9,10,11,12] namely $[Re(CO)_3(N,N)L]^{n+}$ (either L = Cl, Br, I; N,N = 2,2'-bipyridine (bpy); n=0 or L= imidazole (im); N,N = 1,10-phenanthroline (phen); n= 1).

However, the employed protocol—the LVC model in combination with the way in which the model parameters are obtained—might encounter a number of limitations in the case of transition metal complexes. One difficulty often occurring in these complexes is related to the high density of electronic states, which leads to a large number of non-zero coupling terms. This is not very problematic in the Re(I) α-diimine carbonyl halides $[Re(CO)_3(bpy)X]$ (X= Cl, Br, I), as they can be treated by a model involving only sets of pairwise interacting states for symmetry reasons related to the selected electronic states.[5,6] On the contrary, the Re(I) α-diimine carbonyl imidazole [Re(CO)$_3$(phen)(im)]$^+$ complex additionally exhibits interstate coupling between electronic states of the same symmetry that cannot be neglected[7,8] and, in addition, the states cannot be considered strictly to interact pairwise. Another shortcoming of the LVC model is the neglect of second-order terms in the Taylor expansion of the potential energy matrix that might be important if the excited state potentials have very different curvature than the electronic ground state one. This is true in particular when totally symmetric normal modes couple closely lying states of the same symmetry, as we shall discuss in this paper.

The vibronic coupling terms can and are usually obtained by a fit of the computed adiabatic potential energy surfaces[3,4,13], or by analytical formulas based on the computed gradients and hessians of these surfaces[1,14] when the pairwise interacting states approximation is valid. These approaches extract the coupling terms by exploiting information about the energetics of the problem, and discard information contained in the electronic wavefunction.

The purpose of the present work is to propose a new protocol for the computation of linear interstate vibronic coupling constants from the many-electron wavefunctions computed by means of electronic structure methods. The central idea of the protocol is to employ the overlap matrix between the electronic wavefunctions at close-lying geometries as an adiabatic-to-diabatic transformation matrix, such that the LVC parameters can be obtained by means of numerical differentiation. The protocol, which uses ideas borrowed from trajectory surface hopping,[15,16,17] is applicable to wavefunction based methods and time-dependent density functional theory (TDDFT) alike, where in the latter case the wave functions are replaced by auxiliary many-electron wavefunctions.[18,16]

The first two Sections of the paper are devoted to the theory and to the computational methodology. In the third Section, the new technique is applied to two prototype molecules, namely [Re(CO)$_3$(phen)(im)]$^+$ and [Re(CO)$_3$(bpy)Br] for which several ultrafast luminescence experiments and theoretical studies have been recently reported.[5-12] We focus here on energy and symmetry considerations for the selection of active normal modes and the computation of the linear interstate coupling constants either by means of the standard LVC approach[1] or following the new strategy based on the overlap between auxiliary wavefunctions. The non-adiabatic quantum dynamics is performed using coupling constants extracted from both approaches. The results agree well, despite some differences in the numerical values of the coupling constants which are discussed.

# THEORY

*Vibronic coupling theory*

Vibronic coupling theory[1] is used to build a model Hamiltonian based on a diabatic representation of the electronic states. The diabatic Hamiltonian describing $n_{el}$ coupled electronic states is written as

$$\boldsymbol{H}(\boldsymbol{Q}) = \big(T_N + V_0(\boldsymbol{Q})\big)\mathbb{I} + \boldsymbol{W}(\boldsymbol{Q}) \qquad (1)$$

where $T_N$ is the kinetic energy operator, $V_0(\boldsymbol{Q})$ is the potential energy of some reference electronic state, $\mathbb{I}$ is the $n_{el} \times n_{el}$ identity matrix and $\boldsymbol{W}(\boldsymbol{Q})$ the coupling matrix. $\boldsymbol{Q}$ collects the nuclear degrees of freedom. The adiabatic potential energy surfaces $V_n(\boldsymbol{Q})$ are provided as the eigenvalues of $V_0(\boldsymbol{Q})\mathbb{I} + \boldsymbol{W}(\boldsymbol{Q})$.

The reference potential $V_0(\boldsymbol{Q})$ is in general not restricted to any particular form. In typical applications however, we consider photoexcitation from the ground electronic state to electronically excited state(s). In that case, and as will be in our applications to be discussed below, the reference potential $V_0$ is often described using the harmonic approximation for the ground state, written in terms of mass- and frequency-weighted (dimensionless) normal coordinates $Q_i$. We then have[1,3,4]

$$T_N + V_0(\boldsymbol{Q}) = \sum_i \frac{\omega_i}{2}\left(-\frac{\partial^2}{\partial Q_i^2} + Q_i^2\right) \qquad (2)$$

with $\omega_i$ the harmonic frequency of mode $i$. Vibronic-coupling effects arise from the mixing between electronic states along nuclear displacements and become significant at (near)-degeneracy critical geometries such as conical intersections. The elements of the coupling matrix $\boldsymbol{W}(\boldsymbol{Q})$ represent the changes in the excited state potentials with respect to $V_0(\boldsymbol{Q})$ and vary smoothly as a function of nuclear displacements. They are expanded in Taylor series around the reference geometry (here taken to be the Franck-Condon point, $\boldsymbol{Q}=0$):

$$W_{nn}(\boldsymbol{Q}) = \varepsilon_n + \sum_i \kappa_i^{(n)} Q_i + \frac{1}{2}\sum_{i,j} \gamma_{ij}^{(n)} Q_i Q_j + \cdots \qquad (3)$$

$$W_{nm}(\boldsymbol{Q}) = \sum_i \lambda_i^{(nm)} Q_i + \cdots \qquad n \neq m \qquad (4)$$

where in Eq. (3) $\varepsilon_n$ is the vertical excitation energy for state $n$, $\kappa_i^{(n)}$ is the first-order (linear) intrastate coupling constant along mode $i$ for state $n$, $\gamma_{ij}^{(n)}$ the bilinear intrastate coupling constant for mode $i$ and $j$ for state $n$, etc. In Eq. (4), $\lambda_i^{(n)}$ corresponds to the linear interstate coupling between states $n$ and $m$, for $n{\neq}m$. When only linear terms are included in the expansion of $W(Q)$, we refer to the model as linear vibronic coupling (LVC) model. The vibronic-coupling model has been applied to study numerous organic systems (see, eg, pyrazine[19], butatriene cation[20], benzene cation[21] and its fluoroderivatives,[22,23,24] naphthalene and anthracene cations,[25]…), and also transition metal complexes.[26,27,28,29]

In molecules with symmetry, selection rules strongly restrict the number of non-zero electronic-vibrational coupling constants which largely simplifies the problem of parametrizing these constants. For the linear terms, for a two-state problem, one gets.[1]

$$\Gamma_n \otimes \Gamma_Q \otimes \Gamma_m \supset \Gamma_A \qquad (5)$$

where $\Gamma_n$ and $\Gamma_m$ refer to the electronic state symmetry, $\Gamma_Q$ to the normal mode symmetry and $\Gamma_A$ is the totally symmetric irreducible representation of the symmetry point group of the molecule. For molecules with $C_S$ symmetry for instance -with electronic states of A' and A" symmetry and normal modes of a' and a" symmetry- the active coupling modes ($\lambda_i^{(nm)} \neq 0$) will be the non-totally symmetric ones (a") between electronic states of different symmetry and the totally symmetric ones (a') between electronic states of same symmetry. Eq. (5) also implies that only totally symmetric modes can lead to intrastate coupling $\kappa_i^{(n)}$. A given mode will not couple all pairs of states at first order, unless there is no symmetry at all.

The coupling constants $\kappa_i^{(n)}$ and $\lambda_i^{(nm)}$ can be obtained as the first derivatives of the diabatic coupling matrix elements taken at the reference geometry. However, electronic structure calculations give access to adiabatic potential energy surfaces. The question is then how to extract the vibronic coupling constants from quantum chemical calculations. This question is addressed below.

*Interstate coupling from Hessians*

The vibronic-coupling parameters can be extracted from a fitting of the adiabatic potential energy surfaces.[3] This approach is widely employed, using codes such as VCHAM.[13] When only two electronic states are involved, or if only pairs of states interact, one can use analytical formulas instead.[1,14] They have the advantage of providing a clear picture on the role of the intrastate and interstate vibronic coupling. The coupling constants can be deduced from electronic structure calculations using the first and second derivatives of the adiabatic potential energy surfaces $V_n(\mathbf{Q})$ with respect to $Q_i$ at the ground state equilibrium geometry:

$$\kappa_i^{(n)} = \left.\frac{\partial V_n(\mathbf{Q})}{\partial Q_i}\right|_0 \qquad (6)$$

$$\lambda_i^{(nm)} = \sqrt{\left.\frac{1}{8}\frac{\partial^2 (V_m(\mathbf{Q})-V_n(\mathbf{Q}))^2}{\partial Q_i^2}\right|_0} \qquad (7)$$

The determination of $\kappa_i^{(n)}$ is relatively straightforward from the calculation of the gradients at Franck-Condon, provided analytically by many quantum chemistry methods. However, $\lambda_i^{(nm)}$, which measures the repulsion between states *n* and *m* along the mode *i*, involves the second derivative of the potential-energy surfaces.[1]

The analytical formula Eq. (7) is subject to two approximations: (i) it is a two-state formula that is not valid if more than two states interact, (ii) it assumes that second-order effects described by the terms $\gamma_{ij}^{(n)}$ are negligible. The second point is related to the fact that Eq. (7) involves second derivatives, which are also used to determine the second-order coupling constants $\gamma_{ij}^{(n)}$. To illustrate this problem, we consider the case of two interacting states of the same symmetry along a totally symmetric mode. This situation is unlikely to occur in small systems with high symmetry, but may become important when dealing with a high density of electronic states and low symmetry. Evaluating the sum and difference of the second derivatives at $Q=0$ of the adiabatic potentials one gets:

$$\frac{1}{2}\left(\frac{\partial^2}{\partial Q_i^2}V_m + \frac{\partial^2}{\partial Q_i^2}V_n\right)\bigg|_0 = \frac{\gamma_{ii}^{(m)}+\gamma_{ii}^{(n)}}{2} + \frac{\partial^2}{\partial Q_i^2}V_0\bigg|_0 \quad (8)$$

$$\frac{1}{2}\left(\frac{\partial^2}{\partial Q_i^2}V_m - \frac{\partial^2}{\partial Q_i^2}V_n\right)\bigg|_0 = \frac{\gamma_{ii}^{(m)}-\gamma_{ii}^{(n)}}{2} + \frac{2\lambda_i^{(nm)2}}{\varepsilon_m-\varepsilon_n} \quad (9)$$

Eq. (8) shows the average change in the second derivatives of the two states as compared to the ground-state reference potential. In Eq. (9), we can see that $\gamma_{ii}$ terms and $\lambda_i^{(nm)}$ appear. Since there are three unknowns ($\gamma_{ii}^{(m)}$, $\gamma_{ii}^{(n)}$ and $\lambda_i^{(nm)}$) but only two equations, Eqs. (8) and (9) are not sufficient to extract $\gamma_{ii}^{(m)}$, $\gamma_{ii}^{(n)}$ and $\lambda_i^{(nm)}$. Under the assumption of $\gamma_{ii}$ terms equal to zero (or less restrictive, $\gamma_{ii}^{(m)} = \gamma_{ii}^{(n)}$) then Eq. (9) allows to extract $\lambda_i^{(nm)}$ and is equivalent to Eq. (7). Within the harmonic approximation for $V_0$, $\gamma_{ii}$ terms represent the change in frequency of the excited states with respect to the ground state and would affect the values of $\lambda_i^{(nm)}$ extracted from Eq. (7) given that the changes in shape of the potential is due to both (intrastate) quadratic coupling and (interstate) linear coupling. For a more detailed discussion of the Hessian-based approach and evaluation of the error in $\lambda_i^{(nm)}$ due to $\gamma_{ii}^{(m)} \neq \gamma_{ii}^{(n)}$, see Ref [14].

This example shows that alternative ways to extract interstate vibronic coupling constants may be useful, in particular to extract $\lambda_i^{(nm)}$ terms without evaluating the second derivative of the potentials, but also in order to reduce the number of electronic structure calculations as compared to a fitting procedure.

*Interstate coupling from energies and overlaps*

In the approach discussed above, the two-state analytical formulas or fitting strategies only consider energetic information and ignore the information that is contained in the electronic wavefunctions or the TDDFT response vectors. One alternative is thus to consider how this information can be exploited. This can be achieved through the use of overlaps between wavefunctions computed for different molecular structures. First, it should be pointed out that the $W_{nm}$ matrix elements of the above equations can be expressed in the following form,

$$W_{nm} = \langle \Phi_n | H_{el} | \Phi_m \rangle \quad (10)$$

where $\Phi_n$ is the diabatic electronic wavefunction for state *n* and $H_{el}$ the electronic Hamiltonian.

It follows that the $\lambda_i^{(nm)}$ values can be expressed as

$$\lambda_i^{(nm)} = \left.\frac{\partial W_{nm}(Q)}{\partial Q_i}\right|_0 = \left.\frac{\partial}{\partial Q_i} \langle \Phi_n | H_{el} | \Phi_m \rangle \right|_0 \quad (11)$$

This term is closely related to the non-adiabatic coupling (NAC) vector. A direct evaluation of $\lambda_i^{(nm)}$ values from the NAC vectors or the related interstate coupling vectors has been used extensively by Yarkony and coworkers[30,31] in the context of computations at the multi-reference configuration interaction (MRCI) level[32] and has also been achieved in the context of coupled-cluster computations by Ichino *et al.*[33] and Tajti *et al.*[34] However, this type of approach is hampered by the fact that interstate coupling vectors or NAC vectors have only been implemented for a few electronic structure methods based mainly on wavefunction based approaches. Therefore, we will follow a somewhat different route that does not require computation of these coupling vectors. Instead, a finite difference scheme based on the use of wavefunction overlaps is carried out. This scheme is applicable to wavefunction based methods as well as to TDDFT. In the latter case, the response vector is used to construct auxiliary many-electron wavefunctions (see below). For each normal mode $i$ of interest, a finite displacement of the geometry quantified by $\delta Q_i$ is performed yielding vertical transition energies $E_n(\delta Q_i)$. In addition, the overlap,

$$S_i^{(nm)} = \langle \Psi_n(0) | \Psi_m(\delta Q_i) \rangle \qquad (12)$$

is computed, where $\Psi_n(0) = \Phi_n(0)$ is the wavefunction at the reference geometry (where adiabatic and diabatic states coincide) and $\Psi_m(\delta Q_i)$ is the auxiliary wavefunction of the adiabatic state at the displaced geometry. As a next step, a transformation matrix $\boldsymbol{U}$ is constructed by a Löwdin orthogonalization[35] of the overlap matrix $\boldsymbol{S}$. Following Granucci *et al.*[15] and Plasser *et al.*,[36] the diabatic Hamiltonian at the displaced geometry is obtained as

$$\boldsymbol{W}(\delta Q_i) = \boldsymbol{U} \begin{pmatrix} E_1(\delta Q_i) & \cdots & 0 \\ \vdots & \ddots & \vdots \\ 0 & \cdots & E_{n_{el}}(\delta Q_i) \end{pmatrix} \boldsymbol{U}^T \qquad (13)$$

A similar approach to construct diabatic potentials has been proposed by Neugebauer *et al.*[37,38] in the context of linear response theory within TD-DFT, by exploiting overlap between adiabatic transition densities and applied to the computation of vibronic spectra. Here, we shall use auxiliary wavefunction overlaps instead, see the Section Computational Methods below.

The $\lambda_i^{(nm)}$ values are obtained by a numerical differentiation

$$\lambda_i^{(nm)} = \frac{W_{nm}(\delta Q_i)}{\delta Q_i}. \tag{14}$$

The vibronic coupling model discussed so far allows for a diabatization by ansatz of the potential energy surfaces, which are thus constrained to a specific mathematical form. A generalization of the diabatization can be done by exploiting the so-called regularized diabatic states.[39,40,41] Here, the LVC model is used only to define the adiabatic-to-diabatic mixing angle[1,42]: far from the intersection, the diabatic potentials tend to the adiabatic onces. The overlap technique to extract linear coupling constants can be efficiently exploited in this context as well, leading to more flexibility in the construction of diabatic potentials.

Finally, note that it is also possible to obtain the $\kappa_i^{(n)}$ from the overlaps. This is not discussed here since we employ the traditional approach by considering the gradients of the adiabatic potentials in all cases, see Eq. (6). Indeed, analytical gradients are nowadays easily available for numerous quantum chemistry methods.

## COMPUTATIONAL METHODS

*Auxiliary wavefunction overlaps*

We start by commenting on the auxiliary wavefunction overlaps as shown in Eqs.(12-14). In the case of TD-DFT, approximate auxiliary many-electron wavefunctions[18,16] are constructed in the form

$$\Psi_n(\mathbf{Q}) = \sum_{ja} X_{ja}^{(n)}(\mathbf{Q})\, \phi_j^a(\mathbf{Q}). \qquad (15)$$

Here, $X_{ja}^{(n)}$ is the element of the TD-DFT response vector of state $n$ that pertains to the excitation from the occupied orbital $j$ to the virtual orbital $a$, and $\phi_j^a$ is the corresponding Slater determinant. The auxiliary wavefunction overlap Eq. (12) is computed as a double sum of the form

$$S_i^{(nm)} = \sum_{ja} \sum_{kb} X_{ja}^{(n)}(0) X_{kb}^{(m)}(\delta Q_i) \langle \phi_j^a(0) | \phi_k^b(\delta Q_i) \rangle \qquad (16)$$

where $\langle \phi_j^a(0) | \phi_k^b(\delta Q_i) \rangle$ is the overlap between two Slater determinants expressed with respect to non-orthogonal Kohn-Sham orbitals.

In practice, the computation proceeds by first computing the overlap between the atomic orbitals at the two geometries then using these values together with the molecular orbital (MO) coefficients to compute the MO overlaps. The overlap between the Slater determinants is in turn computed as the determinant of the matrix containing all mutual MO overlaps. The evaluation of Eq. (12) can be very costly as it scales with the square of the size of the CI-vector. However, an efficient algorithm,[43] which allows computing the auxiliary wavefunction overlap at a feasible computational cost by applying a prescreening algorithm and by taking advantage of recurring intermediates, is employed here.

***Model Hamiltonian***

The model Hamiltonian used in the present work has been introduced recently[5-7] and includes SOC and vibronic coupling within the LVC model. The Hamiltonian is given by Eqs. (1-2) and the particular form of the $W(Q)$ matrices for our applications are provided in Appendix A. The SOC constants are taken to be constant at the Franck-Condon values. The $W(Q)$ matrices used in this work account for a total of seventeen electronic states for [Re(CO)$_3$(phen)(im)]$^+$ and eleven states for [Re(CO)$_3$(bpy)(Br)] when considering the triplet's components explicitly.

The intrastate coupling terms, $\kappa^{(n)}$, are extracted from the gradients of the excited states, Eq. (6). The interstate coupling constants, $\lambda^{(n,m)}$, are either obtained from the Hessian of the excited states using Eq. (7) (named Hessian-based approach), or computed from the overlap integrals between the excited-state adiabatic wave-functions, Eq. (14) for $\delta Q_i = 0.1$ (named Overlap-based approach).

The details about the electronic structure data used here are reported elsewhere[44,5,7] and are briefly recall below. Calculations were performed by means of DFT including water or acetonitrile solvent corrections based on a conductor-like screening model (COSMO).[45,46,47] The calculations were performed using the B3LYP functional,[48] the D3 parametrization of Grimme[49] and all-electron triple-ξ Slater-type basis set.[50] The scalar relativistic effects were taken into account within the zeroth-order regular approximation (ZORA).[51] The vertical transition energies were computed within TD-DFT[52,53] at the same level described above under the Tamm-Dancoff approximation (TDA)[54]. The SOCs were computed as matrix elements of the scalar relativistic TD-DFT states.[55,56] The normal modes of the singlet electronic ground state S$_0$ (a$^1$A') are used to build the model multidimensional potential energy surfaces. All calculations were done with the ADF2013 code.[57]

*Wavepacket propagation*

The time-dependent Schrödinger equation for the nuclei is solved by employing the multiconfiguration time-dependent Hartree (MCTDH) method.[58,59,60] Here the multiconfiguration nuclear wavefunction is expressed as a linear combination of sums of Hartree products of time-dependent basis functions, known as single-particle functions (SPF). The wavepacket ansatz adapted to the present non-adiabatic problem corresponds to the multiset formulation. The mode combination, number of primitive basis and SPF used in the simulations are given in Appendix B, Tables 3 and 4. The choice is adapted to the small energy differences between the excited states and to the small displacement of the potentials due to modest $\kappa^{(n)}$ coupling terms. Harmonic-oscillator basis sets were employed. The initial wavepacket corresponds to the harmonic ground vibrational state of the ground electronic state $S_0$, promoted at time zero to the $S_2$ absorbing state (see below). The Heidelberg MCTDH Package is used (version 8.4.13).[61]

# RESULTS AND DISCUSSION

*Energy and symmetry considerations for interstate vibronic coupling between excited states of $[Re(CO)_3(N,N)(L)]^{n+}$*

The role of the different ligands on the properties of the low-lying excited-states of $[Re(CO)_3(N,N)(L)]^{n+}$ complexes has been analyzed in previous work[5-8,44,62] and is briefly recalled here. These complexes are characterized by excited states of different character:

singlet and triplet intra-ligand (IL) states and metal-to-ligand charge transfer (MLCT) states. For the complexes with halides, additional contributions of halide-to-ligand charge transfer (XLCT) character mix with the MLCT states. The previous studies led to two main conclusions: (i) the axial L ligand, being X or im, modulates the absorption energy of the optically active singlet state and (ii) the N,N ligand, either phen or bpy, determines the relative stability of the IL and MLCT states with respect to the optically active singlet state. These conclusions are extracted from TD-DFT data reported for different [Re(CO)$_3$(N,N)(L)]$^{n+}$ complexes, including the two complexes discussed here, [Re(CO)$_3$(bpy)(Br)] and [Re(CO)$_3$(phen)(im)]$^{+}$, for which the results are collected in Table 1. For details about the excited-state electronic structure data we refer to Ref. [44] and Ref. [7] respectively. The optically active singlet state around 400 nm corresponds to the S$_2$ state. The complexes are of C$_s$ symmetry.

**Table 1**. TD-DFT transition energies (eV) associated to the low-lying singlet and triplet states of [Re(CO)$_3$(bpy)(Br)] in acetonitrile[44] and [Re(CO)$_3$(phen)(im)]$^{+}$ in water.[7]

| [Re(CO)$_3$(bpy)(Br)] | | | [Re(CO)$_3$(phen)(im)]$^{+}$ | | |
|---|---|---|---|---|---|
| T$_1$ (A") | MLCT/XLCT | 2.84 | T$_1$ (A") | MLCT | 2.98 |
| T$_2$ (A') | MLCT/XLCT | 2.93 | T$_2$ (A') | MLCT | 3.07 |
| S$_1$ (A") | MLCT/XLCT | 2.96 | S$_1$ (A") | MLCT | 3.11 |
| S$_2$ (A') | MLCT/XLCT | 3.13 | T$_3$ (A") | MLCT/IL | 3.24 |
| T$_3$ (A") | IL | 3.22 | S$_2$ (A') | MLCT | 3.40 |
| T$_4$ (A') | MLCT | 3.34 | T$_4$ (A') | MLCT | 3.42 |
| | | | T$_5$ (A') | MLCT | 3.45 |

In the case of [Re(CO)$_3$(bpy)(Br)], the absorbing S$_2$ state is below T$_3$ in energy and the model is truncated such that T$_4$ is not considered. Within this set of states, three possible interstate vibronic coupling potentially couple different symmetry (but same spin) excited states: S$_1$(A")/S$_2$(A'), T$_1$(A")/T$_2$(A') and T$_2$(A')/T$_3$(A"), along a" modes and one interstate vibronic coupling between same symmetry states, T$_1$(A")/T$_3$(A"), along a' modes. Although potentially active, T$_2$(A')/T$_3$(A") and T$_1$(A")/T$_3$(A") coupling constants are small and are neglected in the model: T$_3$ is only coupled by SOC to the other states. . This brings us to the scenario depicted in Figure 1. In the case of [Re(CO)$_3$(phen)(im)]$^+$, the absorbing S$_2$ state is nearly degenerate with T$_4$ and strongly coupled through SOC[7,60] and thus T$_4$ has to be included in the model. Two additional interstate vibronic couplings arise from the inclusion of T$_4$ in the model, that is T$_3$(A")/T$_4$(A') and T$_2$(A')/T$_4$(A'), as depicted in Figure 2. Note that the S$_2$<T$_3$<T$_4$ excited state manifold of [Re(CO)$_3$(bpy)(Br)] becomes T$_3$<S$_2$<T$_4$ in [Re(CO)$_3$(phen)(im)]$^+$. As a consequence, T$_3$ is ascribed to an intermediate state in [Re(CO)$_3$(phen)(im)]$^+$ actively driving population from S$_2$ to T$_1$, in contrast to the case of [Re(CO)$_3$(bpy)(Br)]. In addition, T$_4$ is coupled to the higher lying T$_5$ state. The impact of this coupling onto the potential energy surfaces and the dynamics shall be discussed below.

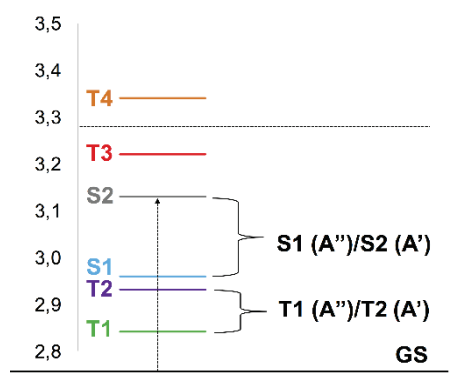

**Figure 1**. Excited states and interstate vibronic coupling considered for [Re(CO)$_3$(bpy)(Br)]. SOC (not shown) additionally couple the states. The ordinate is the exited-state energy in eV.

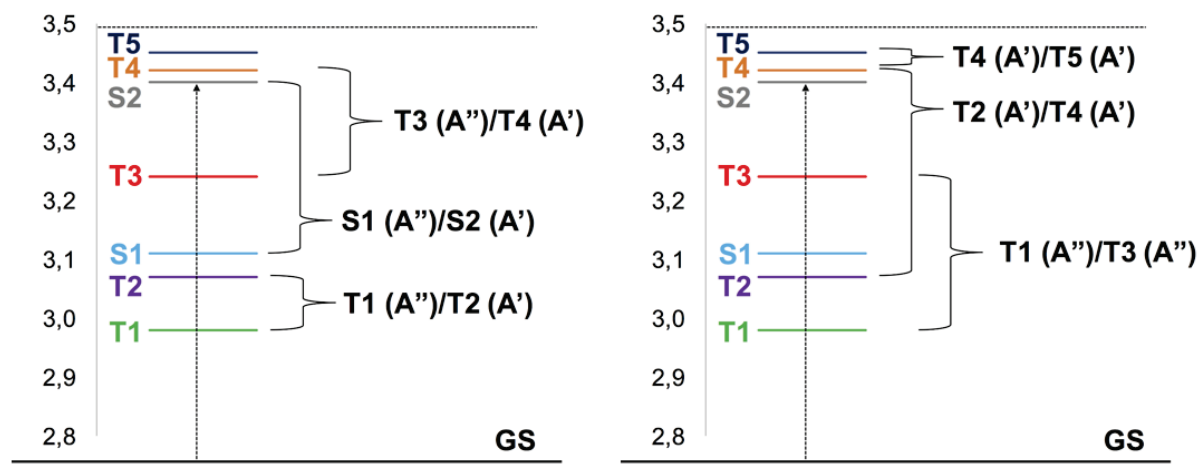

**Figure 2**. Excited states and interstate vibronic coupling between different symmetry states (left) and between same symmetry states (right) considered for [Re(CO)$_3$(phen)(im)]$^+$. SOCs (not shown) additionally couple the states. The ordinate is the exited-state energy in eV.

*Interstate vibronic coupling from hessian model approach and overlap matrix for [Re(CO)$_3$(bpy)(Br)] and [Re(CO)$_3$(phen)(im)]$^+$*

From the 78 and 108 normal modes of [Re(CO)$_3$(bpy)(Br)] and [Re(CO)$_3$(phen)(im)]$^+$, respectively, 14 (12 a' and 2 a") and 15 (12 a' and 3 a") were selected in previous work as the most important ones driving the ultrafast excited state decay from S$_2$ down to the lowest excited state T$_1$.[5-8] This selection was performed by means of analyzing the contribution to the shift in position and energy of the low-lying excited state minima for all the a' modes (obtained from the intrastate couplings $\kappa$). Then, the excited states dynamics were simulated including some interstate vibronic coupling $\lambda$ between different symmetry states (along the a" modes) and between same symmetry states (along the a' modes).[8] As already pointed out, the model used for [Re(CO)$_3$(bpy)(Br)] does not include interstate vibronic coupling between same symmetry states, because the computed coupling constants values are small. The intrastate vibronic couplings were deduced from the gradient of the excited state (see Eq. (6)), and the interstate vibronic coupling $\lambda$ from the Hessians (see Eq. (7)). Herein, the interstate vibronic couplings are also obtained by means of the overlap method. The interstate $\lambda$ values computed by means of both approaches, namely Hessian and Overlap (see previous Sections) are compared in Table 2 for [Re(CO)$_3$(bpy)(Br)] and Figure 3 for [Re(CO)$_3$(phen)(im)]$^+$ for all the coupling modes included in the 14-modes and 15-modes models, respectively.

**Table 2.** Interstate vibronic coupling $\lambda$ (in eV) for the selected a" normal modes (frequency given in cm$^{-1}$) of [Re(CO)$_3$(bpy)(Br)] between different symmetry excited states computed within the Hessian and Overlap approaches.

|  | 95 (cm$^{-1}$) a" | | 486 (cm$^{-1}$) a" | |
| --- | --- | --- | --- | --- |
|  | Hessian | Overlap | Hessian | Overlap |
| S$_1$/S$_2$ | 0.0114 | 0.0016 | 0.0237 | 0.0240 |
| T$_1$/T$_2$ | 0.0086 | 0.0013 | 0.0190 | 0.0268 |

For the bromide complex, only two a" modes are considered, one very low frequency mode at 95 cm$^{-1}$ and one at 486 cm$^{-1}$ corresponding to carbonyl motions[6]. We see in Table 2 that for the mode at 95cm$^{-1}$ the $\lambda$ values extracted from the overlap method are significantly smaller than those obtained from Eq. (7). We attribute this difference to the fact that for such a low frequency mode, the importance of second-order terms with respect to the harmonic approximation can be substantial. As a consequence, the coupling $\lambda$ extracted from the Hessians is increased since it artificially incorporates some second-order contributions, see Eqs. (8)-(9) and the discussion in the Theory section). For the mode at 486 cm$^{-1}$, the $\lambda^{(S_1S_2)}$ values are almost exactly the same because second-order contributions and coupling to higher lying state ($S_3$) are negligible. For $\lambda^{(T_1 T_2)}$ the agreement is less good probably due to some additional indirect coupling to $T_3$ which is not included in the Hessian-based approach. We shall see below that the quantum dynamics using both sets of parameters nearly coincide.

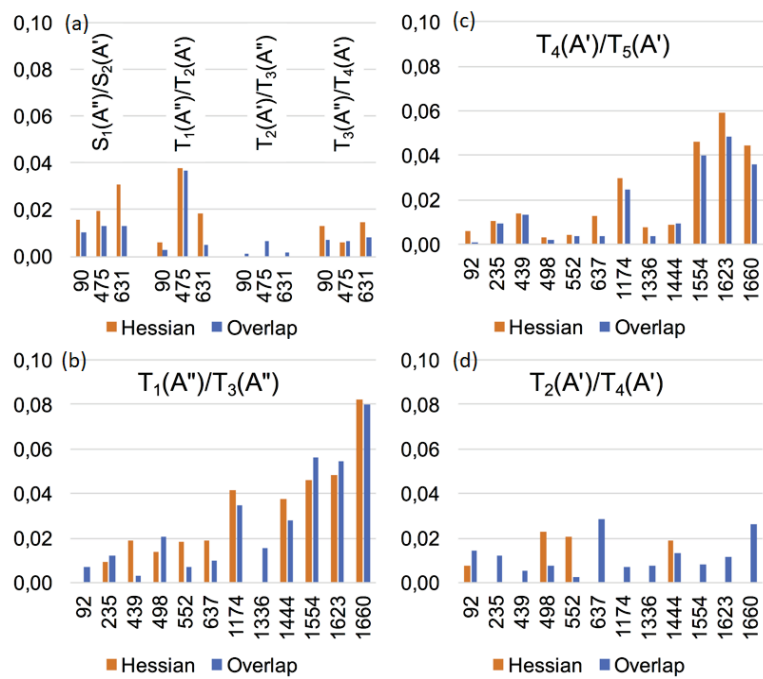

**Figure 3.** Interstate vibronic coupling λ (in eV) for the selected a' and a" normal modes (frequency given in cm$^{-1}$) of $[Re(CO)_3(phen)(im)]^+$ between same symmetry and different symmetry excited states, respectively, computed within the Hessian and Overlap approaches.

The interstate coupling constants between different symmetry states for three a" modes in the case of the $[Re(CO)_3(phen)(im)]^+$ complex are given in Figure 3a. The values obtained from the Hessian-based approach and the overlap approach are rather similar, the former being generally slightly larger than the latter, in particular for $\lambda^{(S_1 S_2)}$. Notice that within the Hessian-based approach the computed $\lambda^{(T_2 T_3)}$ values are zero while those computed with the overlap method are small, less than 0.01 eV, and shall not affect the dynamics (see below).

The three other panels of Figure 3 represent the $\lambda$ values for same-symmetry interacting states along all 12 a' normal modes. For the $T_1(A")/T_3(A")$ and $T_4(A')/T_5(A')$ cases the agreement between the two approaches is good. The only discrepancy is for $\lambda^{(T_1 T_3)}$ along two modes at 92 cm$^{-1}$ and 1336 cm$^{-1}$ for which the Hessian-based values are zero as opposed to the overlap approach. We believe that this is again due to the non-negligible quadratic coupling $\gamma_{ii}^{(n)}$ which makes the effective frequency in $T_3$ smaller or almost equal than that of $T_1$ leading to $\lambda^{(T_1 T_3)} = 0$ in the Hessian based approach (see Appendix C, Figure 7). Indeed, as previously discussed, the $\lambda^{(nm)}$ coupling constant is affected by the quadratic coupling in this approach. In this context, the evaluation of $\lambda^{(nm)}$ through the overlap integral between the adiabatic wave functions provides a Hessian-independent estimation of $\lambda^{(nm)}$ that would allow, in turn, to unequivocally extract $\gamma_{ii}^{(n)}$.

Finally, we inspect the results for the $T_2(A')/T_4(A')$ coupling (Figure 3d). Using the overlap approach, all 12 a' modes exhibit non-vanishing coupling constants, while only 4 modes do so in the Hessian-based approach. This is an artifact of the pairwise interacting state approximation done in Eq. (7). Indeed, we have seen that $T_4$ interacts with $T_5$. As a consequence of this repulsion, the curvature of $T_4$ is lowered. In turn, when computing the $T_2/T_4$ coupling this change in curvature may lead to a zero or negative value of the (square of the) interstate coupling constant (see Eq. (9) in the case $\gamma_{ii}^{(n)} = \gamma_{ii}^{(m)}$), resulting in $\lambda^{(nm)} = 0$. This corresponds to a three-state problem and is illustrated by the $T_2/T_4/T_5$ coupling in [Re(CO)$_3$(phen)(im)]$^+$ along some particular a' normal modes. The *ab initio* adiabatic potentials along the modes at 1554, 1623 and 1660 cm$^{-1}$ are compared to the "SO free" adiabatic potentials obtained within the LVC model in Figure 4. It can be seen that the proper shape of the $T_4$ potential is recovered only when the $T_4/T_5$ coupling is considered. Note that $T_3$ model potentials (orange curve) deviate from the *ab initio* potentials at Q < -1 and Q > 1 along the modes at 1623 cm$^{-1}$ and 1660 cm$^{-1}$, respectively, also because of coupling with higher excited states which are not included in the present study.

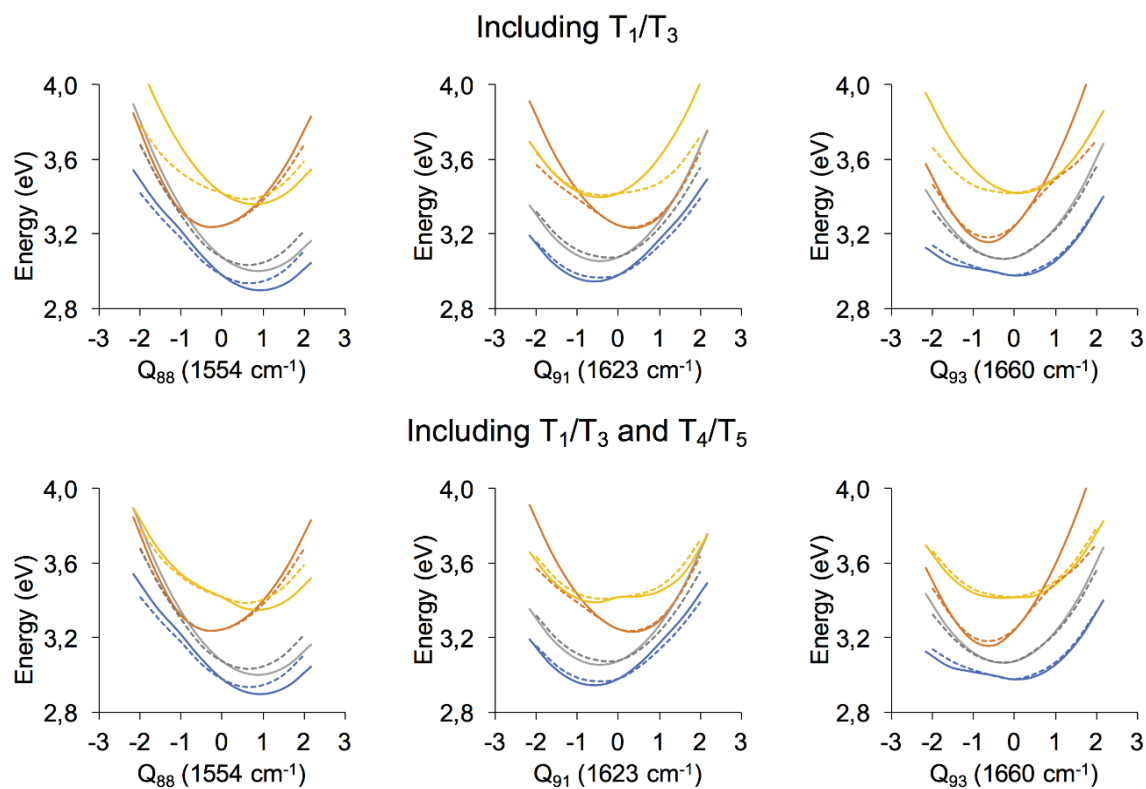

**Figure 4**. Comparison of the model potentials within the triplet manifold of [Re(CO)$_3$(phen)(im)]$^+$ with the *ab initio* "spin-orbit free" adiabatic potentials along modes at 1554 (Q$_{88}$), 1623 (Q$_{91}$) and 1660 (Q$_{93}$) cm$^{-1}$ when including or not T$_4$/T$_5$ vibronic coupling. Color code: T$_1$ blue, T$_2$ grey, T$_3$ orange, T$_4$ yellow (T$_5$ not shown). The thin lines correspond to the Hessian-based model approach and the dotted lines to the *ab initio* data.

Overall, when more than two excited states are vibronically coupled, the pairwise interacting state approach based on the Hessian difference at FC can lead to underestimated $\lambda$ couplings or even the neglect of some of them (as for T$_2$/T$_4$). In this context, the overlap between the excited state adiabatic auxiliary wavefunctions at displaced geometries allows to numerically extract all λ couplings at once including cooperative effects of multiple coupled states. Despite this, the pairwise model approach provides a qualitatively correct representation of the spin-orbit free adiabatic potentials used for the 12 a' modes considered for the $[Re(CO)_3(phen)(im)]^+$ complex, which are compared in Appendix C, Figure 7.

Overall, we believe that this protocol will be highly valuable to study complex systems for which non-zero $\gamma_{ii}^{(n)}$ are important, and/or $\lambda_i^{(n,m)}$ beyond the pairwise interacting state approach must be considered. The consequences on the ultrafast excited state dynamics are discussed in the next section.

## *Quantum dynamics of $[Re(CO)_3(bpy)(Br)]$ and $[Re(CO)_3(phen)(im)]^+$*

In order to assess the role of $T_5$ in the excited state dynamics of the $[Re(CO)_3(phen)(im)]^+$ complex, quantum dynamics by means of wave-packet propagations were performed using the 15-mode model including $T_4/T_5$ interstate vibronic coupling. The evolution of the diabatic electronic populations as a function of time is shown in Figure 5a. All the interstate vibronic coupling constants used in this case were computed within the Hessian-based approach (Figure 3) and thus the results are directly comparable with the previous ones reported in Ref [8] where the same data was used but excluding $T_4/T_5$ couplings. The dynamics is characterized by an ultrafast decay of $S_2$, accompanied by a fast temporary population of the higher-lying $T_4$ and $T_5$ states during a few tens of fs and followed by population transfer to $T_3$ and $T_1$. Then, the $T_3$ population decays to $T_1$ on a longer time-scale. The results presented in Ref 8 where $T_5$ is neglected (thus ignoring the $T_4/T_5$ coupling and its effect in the potential surface shape of $T_4$ as discussed previously) show that $T_4$ plays alone the role of both $T_4$ and $T_5$. We conclude that, while the inclusion of $T_5$ and $T_4/T_5$ coupling in the model is necessary in order to obtain a realistic representation of the $T_4$ potential along some a' modes, it does not modify the ultrafast excited state decay of $S_2$ and neither the population of $T_1$ in this case. This is encouraging since one has to truncate the number of states included in the model, and this truncation can be done without affecting the population transfer to the low-lying states.

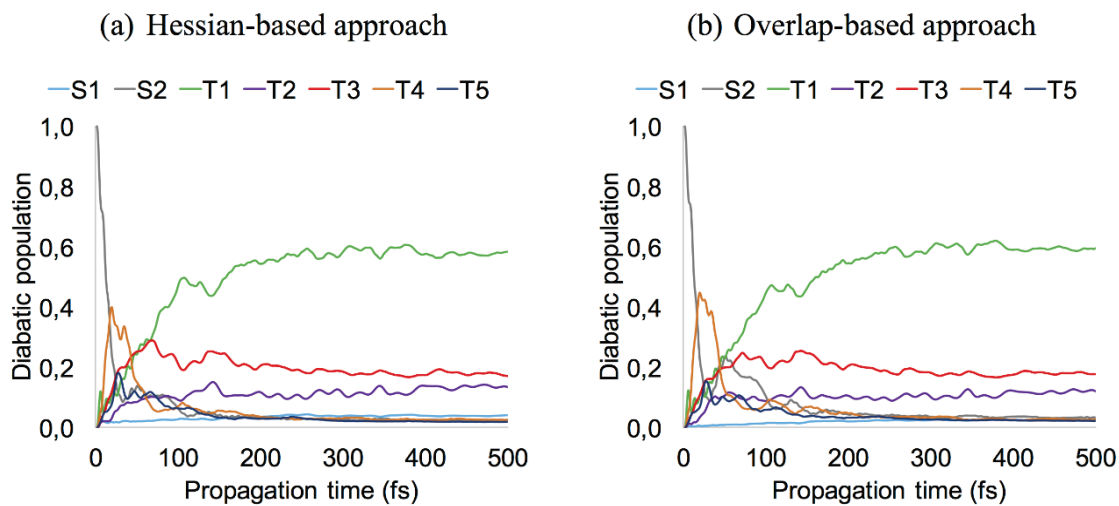

**Figure 5**. Diabatic electronic populations of the low-lying excited states of [Re(CO)$_3$(phen)(im)]$^+$ as a function of time using a 15-modes model (a) considering the interstate vibronic couplings obtained via the Hessian-based approach. (b) considering the interstate vibronic couplings obtained via the Overlap-based approach. Triplet-component contributions are summed up.

After comparing our previous results to the new ones, we now turn to the comparison of the dynamics of the [Re(CO)$_3$(phen)(im)]$^+$ complex obtained using the coupling constants extracted from the Hessian-based approach and the overlap approach

When substituting these values of $\lambda$ by the values computed by means of the overlap approach (Tables 2 and Figure 3) we can see that the small differences discussed in the former section do not influence drastically the ultra-fast non-adiabatic dynamics, both for [Re(CO)$_3$(phen)(im)]$^+$ (compare Figure 5a vs. Figure 5b) and [Re(CO)$_3$(bpy)(Br)] (Figure 6a vs. 6b) complexes. For the two complexes the model is robust against small modifications of the coupling constants and unless very precise results are needed –in which case the use of the LVC model is questionable anyway- it allows to simulate intricate ultrafast decay processes involving many electronic states.

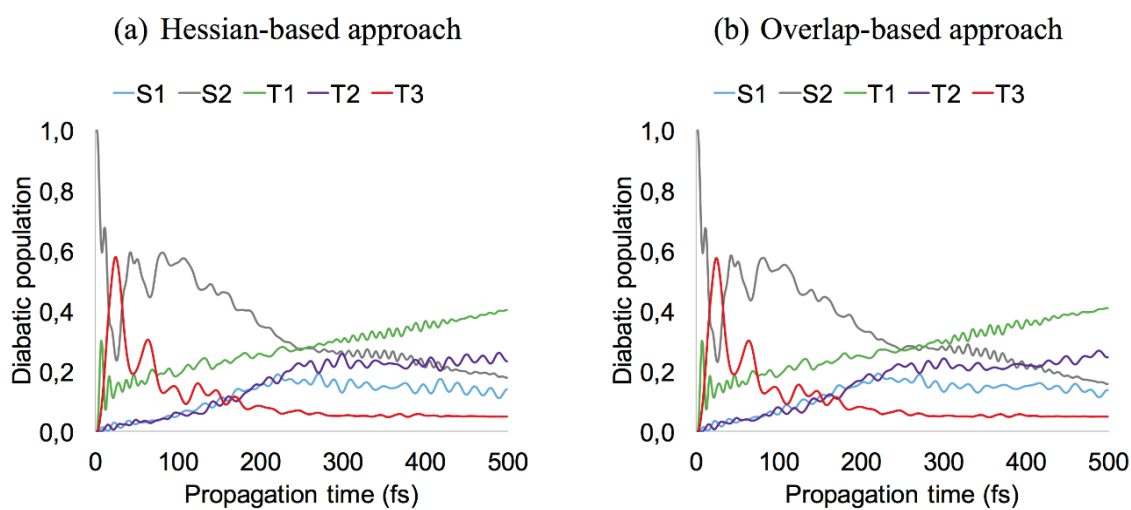

**Figure 6**. Diabatic electronic populations of the low-lying excited states of [Re(CO)$_3$(bpy)(Br)] as a function of time using a 14-modes model:[6] (a) considering the interstate vibronic couplings obtained via the Hessian-based approach. (b) considering the interstate vibronic couplings obtained via the Overlap-based approach. Triplet-component contributions are summed up. Panel (a) reprinted (adapted) from Ref [6]. Copyright (2016) American Chemical Society.

## Conclusions

The construction and use of model vibronic Hamiltonians is very useful in studying the quantum dynamics in the coupled manifold of electronically excited states. The pertinence of the model relies on the adequacy between the shapes of the actual potential energy surfaces with respect to those constructed *by ansatz* in the model. The coupling constants entering the model are extracted from electronic structure data, most often using only the energies, gradients, etc of the excited states. There computation might become quite involved, in particular when many electronic states interact in molecular system of low or even no symmetry.

Here we propose an alternative approach that makes use of information about the vibronic coupling which is embedded in the electronic wavefunctions. More specifically, the vibronic coupling constants are extracted from the overlap of wavefunctions at displaced geometries along the normal modes of interest. While being general, we employ here this strategy in the context of density functional theory, through the use of auxiliary wavefunctions built from the TD-DFT response vector.

The values of the coupling constants obtained from a well known pairwise state-interaction picture and those obtained by the proposed overlap protocol are compared in the case of two prototype rhenium complexes. The values agree qualitatively well and almost quantitatively well for some states and for some modes. The discrepancy, when present, comes from the interaction with more than two states on the one hand, and the inability of the pairwise model to disentangle linear coupling terms from possible contributions of second-order terms. The quantum dynamics performed using the two sets of values for the two complexes show however very little differences as far as the electronic population dynamics is concerned. The model is robust against minor changes in the coupling constants.

We believe that the proposed method is of general interest for computing vibronic coupling constants, and specifically when dealing with complex systems involving a high density of states and low symmetry.

## Acknowledgement

We thank Prof. L. González for fruitful discussions. MF, CD and EG acknowledge funding from the Agence Nationale de la Recherche within projects ANR-10-LABX-0026 (Chimie des Systèmes Complexes) and ANR-15-CE29-0027-01 (DeNeTheor) as well as the Frontier Research in Chemistry Foundation, Strasbourg. The calculations have been performed at the High Performance Computer Centre (HPC), University of Strasbourg, and on the cluster of the Laboratoire de Chimie Quantique (CNRS / University of Strasbourg). F.P and S.M. acknowledge funding from the Austrian Science Fund (FWF) within project I2883 (DeNeTheor) and the University of Vienna, as well as the Vienna Scientific Cluster (VSC) for generous allocation of computing time. Support from the COST action CM1305 (ECostBio) is acknowledged.

**APPENDIX A**

The **W** matrix used for [Re(CO)$_3$(phen)(im)]$^+$ when including the two lowest singlets and five lowest triplets read as follows, where the star stands for the conjugate transpose:

$$\mathbf{W} = \begin{pmatrix} \mathbf{W}^{T1,T1} & \mathbf{W}^{T1,T2} & \mathbf{W}^{T1,S1} & \mathbf{W}^{T1,T3} & \mathbf{W}^{T1,S2} & \mathbf{W}^{T1,T4} & \mathbf{W}^{T1,T5} \\ \mathbf{W}^{*T1,T2} & \mathbf{W}^{T2,T2} & \mathbf{W}^{T2,S1} & \mathbf{W}^{T2,T3} & \mathbf{W}^{T2,S2} & \mathbf{W}^{T2,T4} & \mathbf{W}^{T2,T5} \\ \mathbf{W}^{*T1,S1} & \mathbf{W}^{*T2,S1} & \mathbf{W}^{S1,S1} & \mathbf{W}^{S1,T3} & \mathbf{W}^{S1,S2} & \mathbf{W}^{S1,T4} & \mathbf{W}^{S1,T5} \\ \mathbf{W}^{*T1,T3} & \mathbf{W}^{*T2,T3} & \mathbf{W}^{*S1,T3} & \mathbf{W}^{T3,T3} & \mathbf{W}^{T3,S2} & \mathbf{W}^{T3,T4} & \mathbf{W}^{T3,T5} \\ \mathbf{W}^{*T1,S2} & \mathbf{W}^{*T2,S2} & \mathbf{W}^{*S1,S2} & \mathbf{W}^{*T3,S2} & \mathbf{W}^{S2,S2} & \mathbf{W}^{S2,T4} & \mathbf{W}^{S2,T5} \\ \mathbf{W}^{*T1,T4} & \mathbf{W}^{*T2,T4} & \mathbf{W}^{*S1,T4} & \mathbf{W}^{*T3,T4} & \mathbf{W}^{*S2,T4} & \mathbf{W}^{T4,T4} & \mathbf{W}^{T4,T5} \\ \mathbf{W}^{*T1,T5} & \mathbf{W}^{*T2,T5} & \mathbf{W}^{*S1,T5} & \mathbf{W}^{*T3,T5} & \mathbf{W}^{*S2,T5} & \mathbf{W}^{*T4,T5} & \mathbf{W}^{T5,T5} \end{pmatrix}$$

Note that we explicitly consider the triplet's components, yielding a seventeen states **W** matrix. The different sub-matrices are defined as follows:

$$\mathbf{W}^{n,n} = \varepsilon_n + \sum_{i \in a\prime} \kappa_i^{(n)} Q_i$$

$$\mathbf{W}^{S1,S2} = \sum_{j \in a"} \lambda_j^{(S1,S2)} Q_j$$

$$\mathbf{W}^{Sn(A\prime),Tm(A\prime)} = \mathbf{W}^{Sn(A"),Tm(A")} = \begin{pmatrix} 0; & \eta_{Sn,Tm}; & 0 \end{pmatrix}$$

$$\mathbf{W}^{Tm(A\prime),Sn(A\prime)} = \mathbf{W}^{Tm(A"),Sn(A")} = \begin{pmatrix} 0 \\ \eta_{Sn,Tm} \\ 0 \end{pmatrix}$$

$$\mathbf{W}^{Sn(A\prime),Tm(A")} = \mathbf{W}^{Sn(A"),Tm(A\prime)} = \begin{pmatrix} \eta^*_{Sn,Tm}; & 0; & \eta_{Sn,Tm} \end{pmatrix}$$

$$\mathbf{W}^{\text{Tm}(A'),\text{Sn}(A'')} = \mathbf{W}^{\text{Tm}(A''),\text{Sn}(A')} = \begin{pmatrix} \eta^*_{\text{Sn,Tm}} \\ 0 \\ \eta_{\text{Sn,Tm}} \end{pmatrix}$$

$$\mathbf{W}^{\text{Tn}(A'),\text{Tm}(A'')} = \mathbf{W}^{\text{Tn}(A''),\text{Tm}(A')} = \begin{pmatrix} \sum_{j \in a''} \lambda_j^{(\text{Tn,Tm})} Q_j & \eta_{\text{Tn,Tm}} & 0 \\ -\eta^*_{\text{Tn,Tm}} & \sum_{j \in a''} \lambda_j^{(\text{Tn,Tm})} Q_j & \eta_{\text{Tn,Tm}} \\ 0 & -\eta^*_{\text{Tn,Tm}} & \sum_{j \in a''} \lambda_j^{(\text{Tn,Tm})} Q_j \end{pmatrix}$$

$$\mathbf{W}^{\text{Tn}(A''),\text{Tm}(A'')} = \mathbf{W}^{\text{Tn}(A'),\text{Tm}(A')}$$

$$= \begin{pmatrix} \sum_{i \in a'} \lambda_i^{(\text{Tn,Tm})} Q_i + \eta_{\text{Tn,Tm}} & 0 & 0 \\ 0 & \sum_{i \in a'} \lambda_i^{(\text{Tn,Tm})} Q_i & 0 \\ 0 & 0 & \sum_{i \in a'} \lambda_i^{(\text{Tn,Tm})} Q_i + \eta^*_{\text{Tn,Tm}} \end{pmatrix}$$

where the SOC $\eta_{n,m}$ constants (complex-valued) are obtained from electronic structure calculations at FC. The **W** matrix used for [Re(CO)$_3$(bpy)(Br)] when including the two lowest singlets and three lowest triplets has 11x11 terms analogous to the ones presented above.

**APPENDIX B**

**Table 3.** Number of basis functions for the primitive basis as well as for the time-dependent (SPF) basis used in the MCTDH calculation for the lowest 17 electronic states of $[Re(CO)_3(phen)(im)]^+$ using the 15-modes model Hamiltonian described in the Results section of the main text.

| Modes | Primitive basis | SPF basis |
|---|---|---|
| $(Q_9, Q_{22}, Q_{27})$ | (23, 19, 15) | (11,11,17,11,17,11,11,11,11,11,11,9,9,9,9,9,9) |
| $(Q_8, Q_{24}, Q^{eff}_{36-37})$ | (13, 13, 13) | (8,8,8,8,8,8,8,8,8,8,8,7,7,7,7,7,7) |
| $(Q_{38}, Q_{70}, Q_{81})$ | (17, 17, 17) | (8,8,13,8,13,8,8,8,8,8,8,7,7,7,7,7,7) |
| $(Q_{88}, Q_{91}, Q_{93})$ | (17, 17, 17) | (8,8,11,8,11,8,8,8,8,8,8,7,7,7,7,7,7) |
| $(Q_{18}, Q_{32}, Q_{77})$ | (21, 19, 19) | (9,9,9,9,9,9,9,9,9,9,9,7,7,7,7,7,7) |

**Table 4.** Number of basis functions for the primitive basis as well as for the time-dependent (SPF) basis used in the MCTDH calculation for the lowest 11 electronic states of [Re(CO)$_3$(bpy)(Br)] using the 14-mode model Hamiltonian described in the Results section of the main text.

| Modes | Primitive basis | SPF basis |
| --- | --- | --- |
| ($Q_8$, $Q_{23}$, $Q_{58}$) | (17, 15, 11) | (7,7,7,7,7,7,7,7,7,7,7) |
| ($Q_7$, $Q_{30}$, $Q_{64}$) | (17, 13, 11) | (7,7,7,7,7,7,7,7,7,7,7) |
| ($Q_2$, $Q_{11}$) | (31, 17) | (9,9,9,9,9,9,9,9,7,7,7) |
| ($Q_{12}$, $Q_{13}$, $Q_{16}$) | (17, 17, 17) | (7,7,7,7,7,7,7,7,7,7,7) |
| ($Q_{24}$, $Q_{47}$, $Q_{63}$) | (15, 11, 11) | (7,7,7,7,7,7,7,7,7,7,7) |

# APPENDIX C

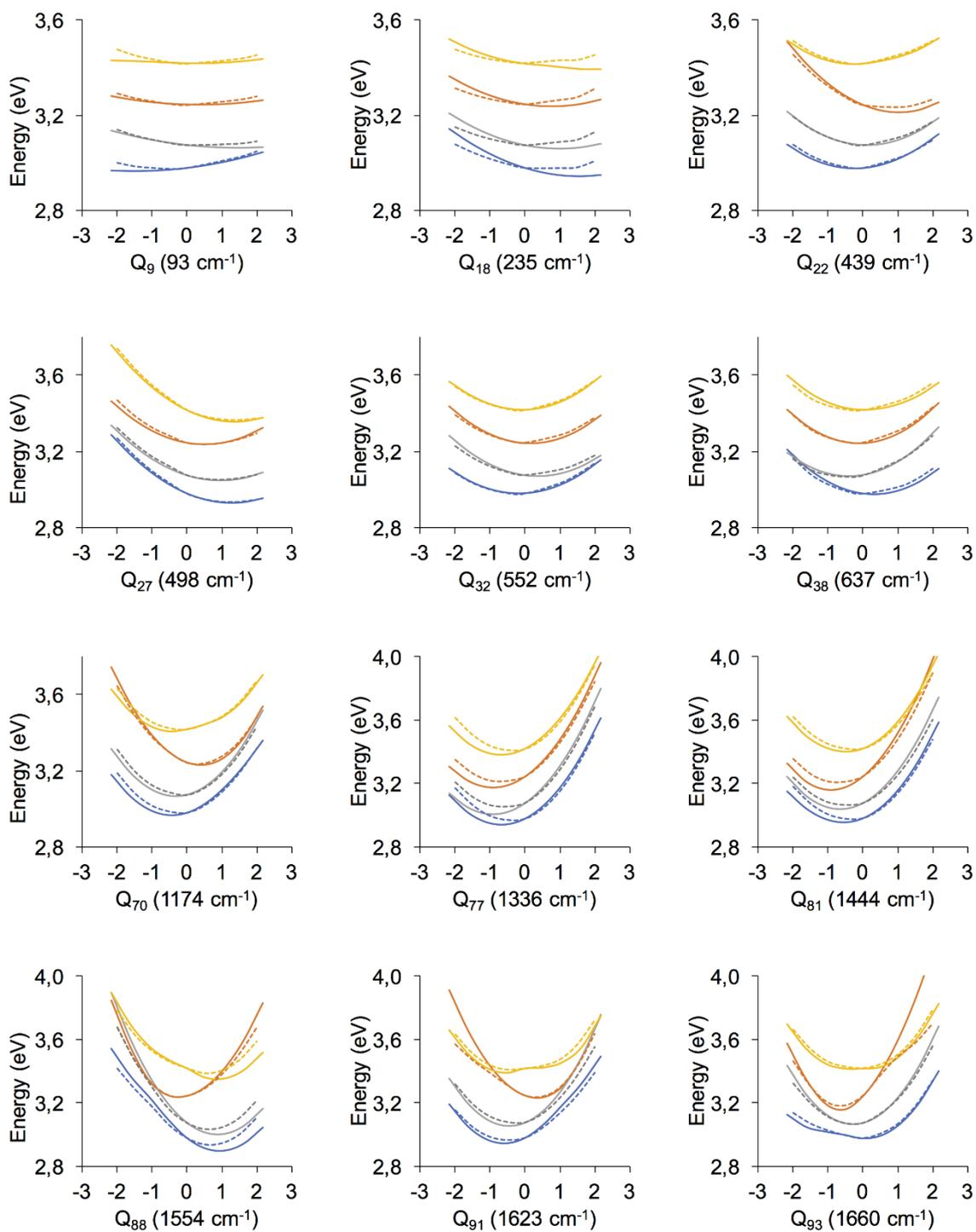

**Figure 7**. Comparison of the model potentials with the *ab initio* "spin-orbit free" adiabatic potentials of the low-lying triplets of [Re(CO)$_3$(im)(phen)]$^+$ along the 12 a' normal modes selected. Color code: T$_1$ blue, T$_2$ grey, T$_3$ orange, T$_4$ yellow. The lines correspond to the model and the dotted lines to the *ab initio* data. Singlet states are not shown.